\documentclass{article}

\usepackage{microtype}
\usepackage{graphicx}
\usepackage{subfigure}
\usepackage{booktabs} 

\usepackage{hyperref}

\usepackage[T1]{fontenc}
\usepackage{xcolor}
\usepackage{listings}
\usepackage{booktabs}
\usepackage{makecell}
\usepackage{tabularx}
\usepackage{comment}

\usepackage{graphicx,amssymb}

\newcommand{\TP}{\textcolor{green!50!black}{\ensuremath{\checkmark}}}
\newcommand{\TN}{\textcolor{green!50!black}{\ensuremath{\times}}}
\newcommand{\FP}{\textcolor{red}{\ensuremath{\checkmark}}}
\newcommand{\FN}{\textcolor{red}{\ensuremath{\times}}}

\definecolor{codebg}{HTML}{F5F7FA}
\definecolor{codeframe}{HTML}{CBD5E1}
\definecolor{codeblue}{HTML}{2563EB}
\definecolor{codegreen}{HTML}{15803D}
\definecolor{codered}{HTML}{B91C1C}

\usepackage{tikz}
\usetikzlibrary{positioning,arrows.meta}
\definecolor{agentfill}{HTML}{FFF2CC}\definecolor{agentdraw}{HTML}{D6B656}
\definecolor{evalfill}{HTML}{F5F5F5}\definecolor{evaldraw}{HTML}{666666}
\definecolor{kgfill}{HTML}{FFE6CC}\definecolor{kgdraw}{HTML}{D79B00}
\definecolor{bestfill}{HTML}{D5E8D4}\definecolor{bestdraw}{HTML}{82B366}

\lstdefinestyle{cuda}{
    language=C++,
    morekeywords={
        __global__, __device__, __host__, __shared__,
        __constant__, __syncthreads, dim3
    },
    basicstyle=\ttfamily\scriptsize,
    keywordstyle=\color{codeblue}\bfseries,
    commentstyle=\color{codegreen}\itshape,
    stringstyle=\color{codered},
    backgroundcolor=\color{codebg},
    frame=single,
    rulecolor=\color{codeframe},
    framerule=0.6pt,
    framesep=3pt,
    numbers=left,
    numberstyle=\tiny\color{gray},
    numbersep=5pt,
    xleftmargin=13pt,
    framexleftmargin=10pt,
    xrightmargin=3.6pt,
    breaklines=true,
    keepspaces=true,
    columns=fullflexible,
    showstringspaces=false,
    tabsize=4,
    captionpos=b
}

\usepackage[preprint]{mlsys2025-arxiv}

\newcommand{\mytool}{KernelArmor}

\mlsystitlerunning{\mytool: End-to-End Agentic System for Establishing Memory and Thread Safety for CUDA}

\begin{document}

\twocolumn[
\mlsystitle{\mytool: End-to-End Agentic System for Establishing Memory and Thread Safety for CUDA}




\begin{mlsysauthorlist}
\mlsysauthor{Bodhisatwa Chatterjee}{nv}
\mlsysauthor{Mohamed Tarek Ibn Zaid}{nv}
\mlsysauthor{Sana Damani}{nv}
\mlsysauthor{Siva Kumar Sastry Hari}{nv}
\mlsysauthor{Christos Kozyrakis}{nv,stanford}
\end{mlsysauthorlist}

\mlsysaffiliation{nv}{NVIDIA}
\mlsysaffiliation{stanford}{Stanford University}

\mlsyscorrespondingauthor
    {Bodhisatwa Chatterjee}{bchatterjee@nvidia.com}



\mlsyskeywords{}

\vskip 0.3in

\begin{abstract}
Optimized CUDA kernels can pass testing while retaining subtle memory and concurrency bugs. We present \textbf{KernelArmor}, an agentic framework that combines LLM-generated annotations with an extended SMT verifier to establish memory and thread safety under explicit contracts. KernelArmor can reason about kernels with advanced CUDA features, such as tensor cores, packed accesses, warp synchronization, asynchronous copies, and library calls, returning source-level diagnostics to guide verification and repair.
Across 740 instances containing 1,370 production-grade CUDA kernels, it reports potential bugs in 78.6\% of instances and establishes safety for 11.4\%, outperforming the evaluated state-of-the-art tools in detection coverage, identifying 100\% of cuCatch memory-safety bugs and all in-scope cuHadron races.
Integrated into an agentic optimization loop, KernelArmor produces verified kernels in all 77 trajectories while retaining 99\% of the unconstrained geometric-mean speedup.
\end{abstract}
]



\begin{NoHyper}
\printAffiliationsAndNotice{}
\end{NoHyper} 

\section{Introduction}
\label{sec:intro}


Over the past few years, agentic frameworks \cite{wang2026openhands, geng2026caid, dai2026cudaagent, brodsky2026kernelforge} have become ubiquitous in code generation, optimization, and repair, yet establishing trust in their outputs remains challenging because they provide no inherent guarantees of correctness. This challenge is further amplified by the growing volume of agent-generated code, which has shifted the bottleneck from code production to manual review \cite{chen2026bottlenecks}, necessitating code validation approaches that are scalable, automated, and rigorous.


A practical approach for validating agent-generated code is testing; however, its effectiveness largely depends on test coverage, and programs that pass the available tests may still contain defects~\cite{liu2023evalplus}. For instance, widely used tools such as NVIDIA \textit{Compute Sanitizer} \cite{nvidiaComputeSanitizer} can detect memory access errors and shared-memory races during execution, but their coverage is limited by the inputs exercised and the error classes supported. In general, this reflects a broader limitation of testing approaches based on \textit{dynamic analysis}: a clean run does not establish the absence of such errors across untested inputs and execution paths~\cite{betts2012gpuverify}.

Formal verification \cite{blom2014gpgpu} tackles this issue by \textit{proving specified properties} across all behaviors permitted by a program’s model and assumptions, thereby providing stronger assurances. However, adapting formal verification techniques at scale remains challenging since constructing specifications, generating program invariants, and writing proof annotations often require substantial expertise and manual effort~\cite{poesia2024dafnyannotator}. This process can also be error-prone and often requires iterative refinement. This effort is further complicated by proof instability in SMT-based formal verifiers, where small changes can cause substantial slowdowns or verification failures~\cite{zhou2023mariposa}.

Recent work~\cite{yang2025autoverus,tihanyi2023esbmcai,chatterjee2025proofwright,martinez2026kuiperbench,browne2026gpuverification,costa2026aicompiler,gargary2026resolve} has extensively explored the use of AI to automate formal verification techniques. These approaches leverage large language models (LLMs) to propose proof artifacts, which are then checked by existing formal verifiers and proof frameworks \cite{lattuada2023verus,cordeiro2012esbmc,blom2017vercors,martinez2026kuiper,driscoll2026volta}. The verifier feedback is then used to guide subsequent revisions, enabling agents to iteratively refine annotations, proofs, or program implementations.

However, applying these techniques to optimized CUDA code raises two challenges. First, existing GPU verifiers~\cite{betts2012gpuverify,li2012gklee,monteiro2018esbmc,cogumbreiro2021faial} support only a small subset of CUDA. Optimized kernels, whether written by experts or by agents, routinely rely on features outside that subset, such as warp-level shuffles, asynchronous memory copies (\texttt{cp.async}), inline PTX, and tensor-core instructions. A verifier that cannot model these features must either reject the fastest kernels or leave them unchecked. Second, placing a verifier inside an agentic optimization loop changes what it must report. A verdict helps the agent only if it points to the failing safety obligation in the source and separates three cases that call for different actions: a likely defect the agent should fix, a proof the verifier could not complete, and a construct outside the verifier's model. Conflating these cases either lets unsafe code through or steers the agent away from correct optimizations. Addressing both challenges requires agentic verification that covers the CUDA features optimized kernels actually use and returns explicit, checkable guarantees and feedback the agent can act on. 

We present \textit{\mytool{}}, an agentic framework for automated memory and thread-safety verification of optimized CUDA kernels. The framework uses in-context learning and reusable knowledge from successful verification attempts to generate and iteratively refine contracts, invariants, and permission annotations for VerCors, an SMT-based deductive verifier for parallel programs~\cite{blom2017vercors}. We extend VerCors' CUDA verification infrastructure to reason about advanced constructs, including packed memory accesses, warp-level synchronization, asynchronous copies, and tensor-core operations, while making their modeling assumptions explicit. When verification fails, \mytool{} provides fine-grained, source-grounded feedback on unresolved safety obligations and potential defects, guiding targeted revisions to annotations or implementations within agentic code-generation loops.


We evaluated \mytool{} on three benchmark suites ($\sim$1,400 kernels) containing production-grade CUDA kernels, identifying potential safety bugs in \textbf{78.6\%} of instances, and establishing memory and thread safety for \textbf{11.4\%}. Compared with state-of-the-art memory and thread safety tools, \mytool{} achieved \textbf{100\%} detection on the cuCatch memory safety suite~\cite{ibnziad2023cucatch} and all in-scope data races in the cuHadron suite~\cite{stephenson2026supercollider}, with no false positives. Finally, integrating \mytool{} into an agentic optimization loop yielded verified kernels in all 77 trajectories while retaining \textbf{99\%} of the geometric-mean speedup achieved without verification.

Overall, this paper makes the following contributions:
\begin{itemize}
    \item We present \mytool{}, an end-to-end agentic system that combines reusable verification knowledge, iterative verifier feedback, and an annotation-integrity guard to establish memory and thread safety in production-grade CUDA programs.
        
    \item We extend the CUDA frontend, memory model, and synchronization semantics of VerCors~\cite{blom2017vercors}, an SMT-based verifier, to support advanced CUDA features, including asynchronous copies, packed accesses, warp-scoped synchronization, dynamic shared memory, and library calls with explicit safety contracts.

    \item We outperform state-of-the-art memory and race detectors and enable verified agentic optimization in the loop while retaining 99\% of the unconstrained geometric-mean speedup.
\end{itemize}


\section{Background \& Motivation}
\label{sec:motiv}

Modern CUDA kernels can contain subtle memory and concurrency bugs that escape off-the-shelf safety tools. In particular, input-dependent global-memory races and accesses beyond a tensor view's logical bounds can evade tools whose checks cover only selected memory spaces or allocation-level bounds.

\subsection{Safety Issues in Modern CUDA Kernels}

Listing \ref{lst:cuda-cache-race} shows an AI-optimized kernel with a
global-memory race when \texttt{cache\_position = [0,0]},
\texttt{batch = 1}, and \texttt{seq\_len = max\_seq\_len = 2}:
corresponding threads in blocks 0 and 1 access the same packed BF16
cache entry. If block 0 reads and clears the entry before block 1 loads
it, block 1 receives zeros, producing an incorrect gradient instead of
the original cache value that the reference gathers for both sequence
positions.

\begin{lstlisting}[style=cuda,
    caption={Duplicate cache positions causing a global-memory race.},
    label={lst:cuda-cache-race}]
__global__ __launch_bounds__(256, 2)
void fused_rope_cache_backward(..) {
// ... compute parameters ...
const int64_t pos = cache_position[seq];
const int64_t cache_lo =
    ((static_cast<int64_t>(batch) * kHeads + head)
      * max_seq_len + pos) * kDim + pair;
// Blocks alias this entry when positions repeat.
const __nv_bfloat162 v_lo_b =
    *reinterpret_cast<const __nv_bfloat162*>(
        grad_value_cache + cache_lo);
*reinterpret_cast<__nv_bfloat162*>(
    grad_value_states + state_lo) = v_lo_b;
// This clear can race with another block's load.
const __nv_bfloat162 zero =
    __floats2bfloat162_rn(0.0f, 0.0f);
*reinterpret_cast<__nv_bfloat162*>(
    grad_value_cache + cache_lo) = zero;
// ..rest of the kernel..
}
\end{lstlisting}

\mytool{} exposes the overlapping addresses by inserting the following
checked separation assertion immediately before the first cache load:
\begin{lstlisting}[
    style=cuda,
    numbers=none,
    xleftmargin=3.6pt,
    framexleftmargin=0pt,
    language={},
    basicstyle=\ttfamily\scriptsize\color{black},
    morekeywords={assert},
    keywordstyle=\color{violet}\bfseries,
    emph={pointer_block_offset},
    emphstyle=\color{codeblue}
]
/*@ assert
    \pointer_block_offset(grad_value_cache) + cache_lo
    != \pointer_block_offset(grad_value_cache)
       + (head * max_seq_len
          + cache_position[1 - blockIdx.x])
         * kDim + pair;
@*/
\end{lstlisting}
The left-hand side denotes the current thread's cache offset; the
right-hand side denotes the corresponding thread's offset in the other
block. For thread 0 in either block, both sides evaluate to zero, so
VerCors rejects the assertion with \texttt{assertFailed:false}.
Together with the subsequent non-atomic stores and the absence of
inter-block synchronization, this establishes a \textit{global-memory race}.
These accesses are aligned and in bounds but their global-memory
read--write conflict is not currently detected by off-the-shelf tools
like Compute Sanitizer.



\begin{lstlisting}[style=cuda,
    caption={Vectorized bias loads for packed BF16 GELU.},
    label={lst:cuda-bias-gelu}]
__global__ void bias_gelu_bth_vec_kernel(..) {    
const uint4* bias_vec =
    reinterpret_cast<const uint4*>(bias);

for (int64_t idx = int64_t(blockIdx.x) * blockDim.x
                   + threadIdx.x;
     idx < total_vecs; idx += stride) {
    int vec_idx = int(idx % kDModelVecs);
    uint4 v = reinterpret_cast<uint4*>(x)[idx];
    // Reads eight BF16 elements.
    uint4 b_v = bias_vec[vec_idx];
    // ... packed BF16 GELU ...
}
\end{lstlisting}

Listing \ref{lst:cuda-bias-gelu} shows an unchecked bias extent in a fused BF16 bias/GELU kernel, where each \texttt{uint4} load consumes eight BF16 elements. For a
one-element bias view, the first load reads seven elements beyond the tensor's logical extent, allowing unrelated values from the backing allocation to enter the computation. The relevant VerCors annotation, which flags this issue, is:
\begin{lstlisting}[
    style=cuda,
    language={},
    numbers=none,
    xleftmargin=3.6pt,
    framexleftmargin=0pt,
    basicstyle=\ttfamily\scriptsize\color{black},
    morekeywords={assert},
    keywordstyle=\color{violet}\bfseries,
    emph={pointer_length},
    emphstyle=\color{codeblue}
]
//@ assert 8 * vec_idx + 7
//         < \pointer_length(bias);
\end{lstlisting}
This assertion requires all eight elements to lie within the view
and fails for this input. If the view resides within a larger
initialized allocation, the accesses remain within that allocation, so allocation checks performed by most off-the-shelf tools cannot detect the
logical overread without additional region metadata.

\subsection{VerCors Verification Tool}

VerCors~\cite{blom2017vercors} is a deductive verifier for concurrent and parallel programs that can establish memory safety, data-race freedom, and functional correctness. It uses permission-based separation logic to specify
program contracts, loop invariants, and memory-access permissions. VerCors
translates annotated programs into Viper's intermediate verification
language ~\cite{muller2016viper} and checks them using Silicon, Viper's symbolic-execution
backend, which relies on the Z3 SMT solver \cite{demoura2008z3}. A program contract consists of \textit{preconditions}, which state the properties and permissions required at entry, and \textit{postconditions}, which describe those guaranteed on return when the preconditions hold.
For GPU kernels, permissions describe per-thread memory ownership and its
transfer at synchronization points. In \mytool{}, VerCors uses
the agent's annotations to prove memory and thread safety under
explicit preconditions and the supported CUDA model.
\section{\mytool{}: End-to-End Agentic System for Establishing Memory \& Thread Safety}
\label{sec:proofwright}

\mytool{} combines LLM-guided annotation generation with formal verification to assess the memory and thread safety of CUDA programs.



\subsection{\mytool{}'s Agentic Verification Workflow}

\begin{figure*}[t]
\centering\includegraphics[width=1.0\textwidth]{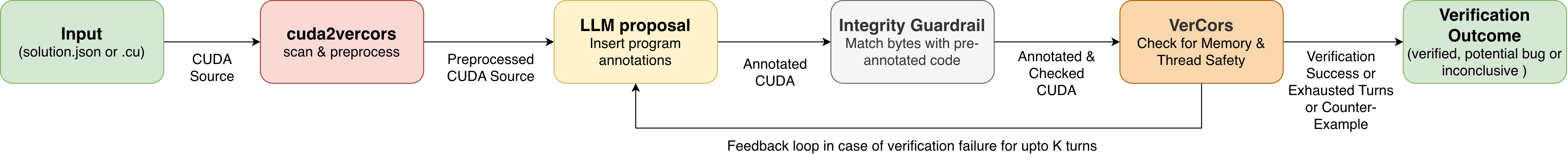}
\caption{\textbf{\mytool's memory- and thread-safety verification workflow}. CUDA inputs are preprocessed for LLM-generated annotation insertion, while an integrity guard ensures byte-for-byte preservation of the preprocessed code after removing annotations. VerCors checks the annotated program and guides repair for up to $K$ turns, yielding a verified, potential-bug, or inconclusive outcome.}
\label{fig:proofwright2.0flow}
\end{figure*}

\mytool{} accepts CUDA code, along with any available host code, either as a \texttt{.cu} source file or within an agent-generated \texttt{.json} file. The input program is first scanned to ensure that all constructs are supported by \mytool{}, then preprocessed by the workflow's frontend into a representation compatible with VerCors. \mytool{} then generates safety annotations, after which an \textit{annotation-integrity guard} ensures that only annotations have been added. The annotated CUDA program is then verified by VerCors, and if required, the annotations are iteratively refined using verifier feedback within a bounded number of attempts. Figure \ref{fig:proofwright2.0flow} illustrates this workflow. The workflow concludes with one of three outcomes: \textit{verified}, \textit{potential bug}, or \textit{inconclusive}. We will now describe each stage of the agentic workflow in detail. 

\textbf{Scanning \& Preprocessing}: \mytool{}'s deterministic (non-agentic) frontend parser, \textit{cuda2vercors}, first scans the input for unsupported constructs, then normalizes CUDA syntax and lowers supported operations, such as inline PTX loads and stores, to calls with explicit safety contracts. If any unsupported constructs are detected, the workflow stops and reports the need for additional preprocessing support, giving users the option to extend \mytool{} offline before rerunning verification.


\textbf{Knowledge-Guided Annotation Generation}: \mytool{}'s LLM proposes preconditions, postconditions, loop invariants, and synchronization annotations for the preprocessed program. The agent also examines host code (if available) and extracts launch metadata to relate kernel preconditions to caller-side allocations, input dimensions, and launch configurations. These preconditions capture the bounds, alignment, and memory permissions required for safe execution, while postconditions specify the permissions preserved on completion. Loop invariants track access permissions and index bounds across iterations, and synchronization annotations describe permission transfers between threads at barriers.

To support efficient generation of safety annotations, \mytool{}'s annotation agent is equipped with a reusable \emph{CUDA Verification Handbook}, which distills VerCors specification concepts into practical annotation patterns, examples, and diagnostic repair strategies. The goal here is to reuse established verification strategies rather than derive them from scratch for each program. Handbook material is included in the agent's prompt to support \textit{in-context learning}, enabling the agent to adapt these strategies to the program's memory accesses, thread ownership, and synchronization requirements. The handbook is updated only when a candidate that demonstrates a novel reusable annotation strategy not already in the handbook is successfully verified.




\textbf{Integrity Guardrail}: To ensure that LLM annotations leave the program unchanged, \mytool{} performs an integrity check which removes annotation comments from the candidate and preprocessed baseline, then checks that the resulting code is byte-for-byte identical. 

\textbf{VerCors Verification}: Finally, the annotated program is checked by VerCors for memory and thread safety under the proposed contracts. On failure, the verifier diagnostic is used to guide annotation repair, and each revised proposal passes through the integrity guard before re-verification. This regeneration-verification loop is bounded by up to $K$ turns, and concludes with one of three outcomes: \emph{verified} if all required checks pass, \emph{potential bug} if validated safety evidence supports a finding, or \emph{inconclusive} otherwise.

\subsection{Extending VerCors' CUDA Frontend, Memory Model and Synchronization Semantics}

We describe \mytool{}’s extensions to VerCors’ frontend, verification encodings, and permission model for optimized CUDA kernels.

\textbf{Specializing CUDA constructs}:
CUDA kernels often use templates and compile-time constants to parameterize
storage sizes and execution paths. \mytool{} first extends VerCors' CUDA frontend
to specialize supported templates and process associated constants, casts,
and device helpers. For example, instantiating \texttt{reduce<128>} makes
parameter-dependent array bounds and compile-time branch choices explicit
before verification. Unsupported constructs are reported as frontend errors.

\textbf{Tracking asynchronous-copy completion}:
Asynchronous copy (\texttt{cp.async}) transfers data from off-chip global memory to on-chip shared memory
without blocking the issuing thread; an explicit wait establishes completion
before that thread consumes the copied data. \mytool{} adds a pending-copy resource that holds the required buffer permissions
until matching completion. Premature destination access or source modification
then fails a permission check. Completing the wrong copy group or buffer pair
fails to match the pending resource, preventing an unrelated wait from
incorrectly authorizing buffer reuse.

\textbf{Checking warp-scoped synchronization}:
Warp-scoped synchronization lets selected lanes within the same warp wait for one another before proceeding. \mytool{} extends VerCors with barriers restricted to the lanes selected by
\texttt{\_\_syncwarp(mask)} within the caller's warp. Participating lanes
must agree on the mask, and permission transfers remain within that group.
An access relying on synchronization with another warp therefore fails its
permission check. Supported shuffles similarly check lane participation
and legal widths, leaving the exchanged values abstract.

\textbf{Modeling shared storage and thread-local state}:
Threads within a CUDA block share dynamic shared memory, but each thread receives its own copy of kernel arguments.
\mytool{} derives shared-array bounds from the launch-provided byte count
and models modified or address-taken arguments as thread-local copies. Thus, $4n$ shared bytes provide space for $n$ floats, while
\texttt{input += stride} updates only the calling thread's cursor. Invalid allocation sizes fail size checks, and dereferences still require
valid bounds and permissions.

\textbf{Preserving pointer arithmetic}:
Pointer arithmetic involves advancing a given pointer in units of the pointed-to type. \mytool{} extends VerCors' cast encoding to preserve the corresponding byte offsets: advancing a BF16 pointer by one moves two bytes, whereas advancing a \texttt{uint4*} by one moves sixteen bytes. Accesses through the cast pointer must still satisfy bounds and permission checks, preventing a change of type from hiding an invalid access to the underlying storage.

\textbf{Checking packed accesses}:
Packed accesses read or write multiple adjacent elements through a single
wider memory operation. \mytool{} lowers supported packed transactions
into scalar accesses while retaining their original alignment requirements.
For instance, a sixteen-byte \texttt{uint4} load over BF16 storage requires eight readable
elements and a sixteen-byte-aligned address. Insufficient bounds, missing
permissions, or misalignment therefore remain proof failures after scalarization.

\textbf{Checking block-barrier participation and ownership}:
A block barrier lets threads within a block wait for one another before
accessing data produced by other threads.
\mytool{} extends barrier encodings to check permission transfers and adds
bounded convergence checks for supported device-helper expansions.
For example, \texttt{\_\_syncthreads()} may redistribute exclusive cell
ownership into read shares, but reaching it only when
\texttt{threadIdx.x == 0} cannot establish participation across a multi-thread
block. Subsequent accesses fail if the barrier contract does not provide
the required permissions.

\textbf{Checking library-call requirements}:
\mytool{} adds typed safety contracts for supported cuBLAS, cuBLASLt, CUTLASS,
WMMA, and PTX operations. These contracts expose buffer extents, layouts,
alignment, and access permissions. Undersized buffers, missing permissions,
or prohibited output overlap fail the call's preconditions. In the current version, the contract assumes the correctness of
the external library implementation.

\textbf{Improving backend permission reasoning}:
\mytool{} patches Silicon, Viper's symbolic-execution verification
backend~\cite{muller2016viper}, to improve permission selection and predicate
matching while retaining the required permission checks. 

Other extensions preserve supported atomic return values and exclusive slot ownership for unit-increment allocation counters, give bounded proof views for selected cluster accesses without changing block-barrier scopes, and preserve or synthesize quantifier triggers for ownership expressions.


\section{Implementation}

We built our extended VerCors fork from source\footnote{https://github.com/utwente-fmt/vercors} and used the NeMo Agent Toolkit~\cite{nvidia_nemo_agent_toolkit_2026} to orchestrate \mytool{}'s workflow. Our extensions comprise approximately 20.6 KLOC of net implementation changes, excluding tests, supporting CUDA 13.1. The preprocessing frontend and annotation agent are implemented in Python 3.13.12, while the VerCors extensions are primarily implemented in Scala 2.13.12.

\section{Evaluation}
\label{sec:eval}

We evaluate \mytool{} along two dimensions: its ability to reason about memory and thread safety in CUDA code produced by agentic and non-agentic approaches, and the impact of integrating it into an agentic code-generation loop. Our experiments address the following research questions:

\begin{itemize}
    \item \textbf{RQ1}: To what extent can \mytool{} reason about memory and thread safety in CUDA kernels that use advanced language and hardware features? What classes of memory and concurrency bugs can it identify?

    \item \textbf{RQ2}: What are the overheads of \mytool{} in terms of runtime, token usage, and estimated inference costs? 

    \item \textbf{RQ3}: How does \mytool{} compare with other state-of-the-art tools that detect CUDA memory and thread safety issues?

    \item \textbf{RQ4}: To what extent does the verification handbook improve \mytool{}'s verification success rate?

    \item \textbf{RQ5}: What is the impact of integrating \mytool{} into an agentic code-generation loop?
\end{itemize}

\textbf{Experimental Setup:} We ran \mytool{} on an Ubuntu server equipped with a 16-core Intel Xeon Gold 6444Y processor (32 hardware threads) and 247\,GiB of system memory. We used GPT-6 Astra for all of the annotation generation and refinement experiments. 

\subsection{Establishing Memory \& Thread Safety in Production-Grade CUDA Kernels}

We evaluate \mytool{} on three suites totaling 740 instances and 1,370 CUDA kernels. In our experiments, an \emph{instance} denotes an individual implementation of a benchmark workload, including its associated host code and potentially multiple CUDA kernels.

\textbf{Evaluation Suite}. We first evaluate \mytool{} on \textbf{SOL-ExecBench} \cite{lin2026solexecbench}, with 64 instances across 46 problems (235 kernels), extracted from production and emerging AI workloads; an \textbf{Optimized CUDA Suite}, with 28 production-grade implementations (70 kernels) covering attention, normalization, convolution, and matrix operations; and \textbf{KernelBench}~\cite{ouyang2025kernelbench}, with 648 generated instances across 1,065 kernels. Collectively, these implementations exercise advanced CUDA features including packed vector and alignment-sensitive BF16 accesses, inline PTX, shared-memory tiling and synchronization, warp-level reductions, atomics, template specialization, WMMA Tensor Core operations, and asynchronous global-to-shared transfers (\texttt{cp.async}). Their execution paths also include FP8/NVFP4 processing, CUDA Graphs, thread-block cluster launches, and cuBLAS/cuBLASLt calls. 

\begin{figure}[!ht]
\centerline{\includegraphics[width=1.0\linewidth]{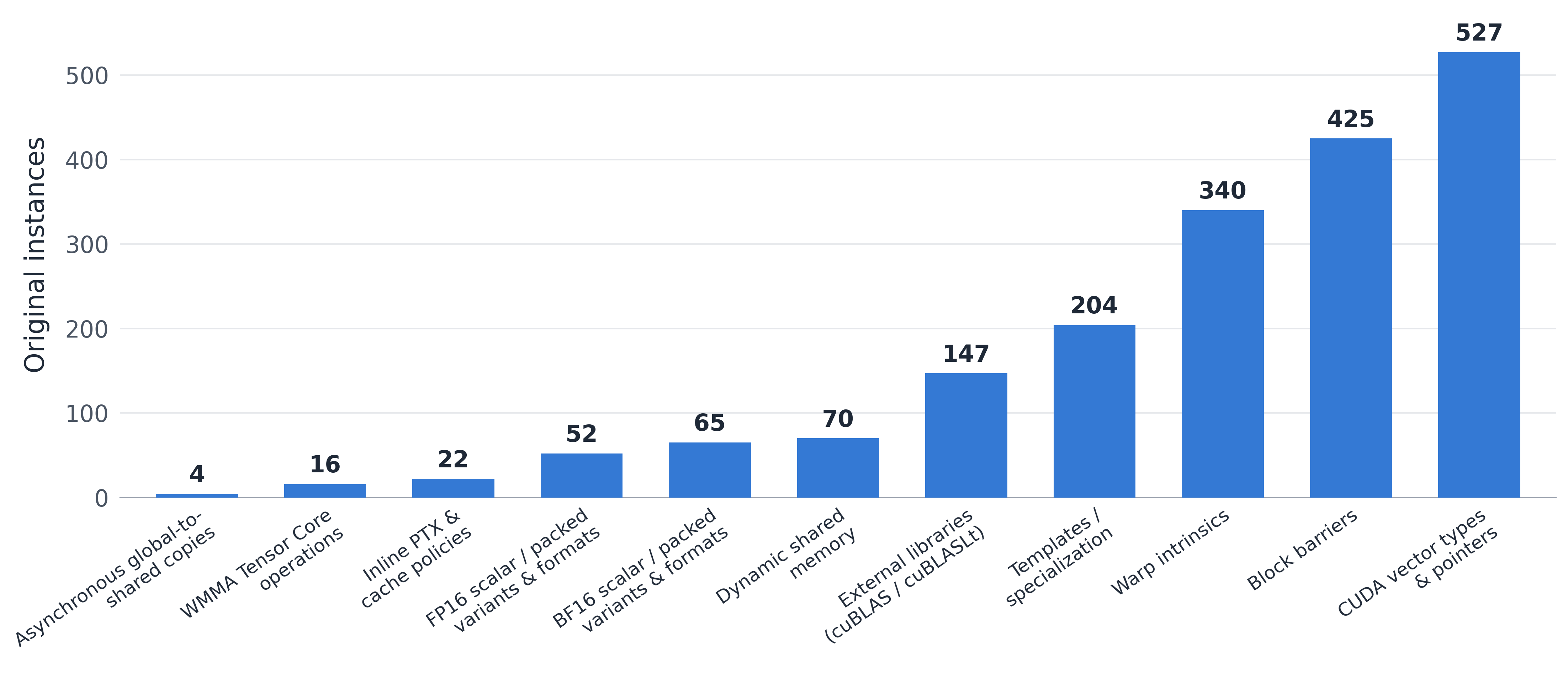}}
\caption{\textbf{CUDA features represented in our evaluation suites}. Bars count instances containing each feature across all three suites.}
\label{cuda-feat}
\end{figure}

\textbf{Verification Outcomes}. Table~\ref{tab:verification-outcomes} illustrates \mytool{}'s verification outcomes across all three evaluation suites. Overall, \mytool{} is able to fully establish memory and thread safety for 84 instances, reports potential findings in 582, and is inconclusive on 74. The high rate of potential bugs highlights the prevalence of safety issues in agent-generated kernels. In each potential bug case, \mytool{} returns a counterexample that triggers the bug. One observation is that these potential bugs also occur in relatively simple workloads (82\% in \textit{activation} workloads). These implementations rely on input-dependent indexing and launch-grid arithmetic, which often lead to overflows and alignment mismatches. Matrix and fused-GEMM workloads account for 75\% of KernelBench's inconclusive outcomes, highlighting some of the library-modeling limitations in \mytool{}.  

\begin{figure}[!ht]
\centerline{\includegraphics[width=1.0\linewidth]{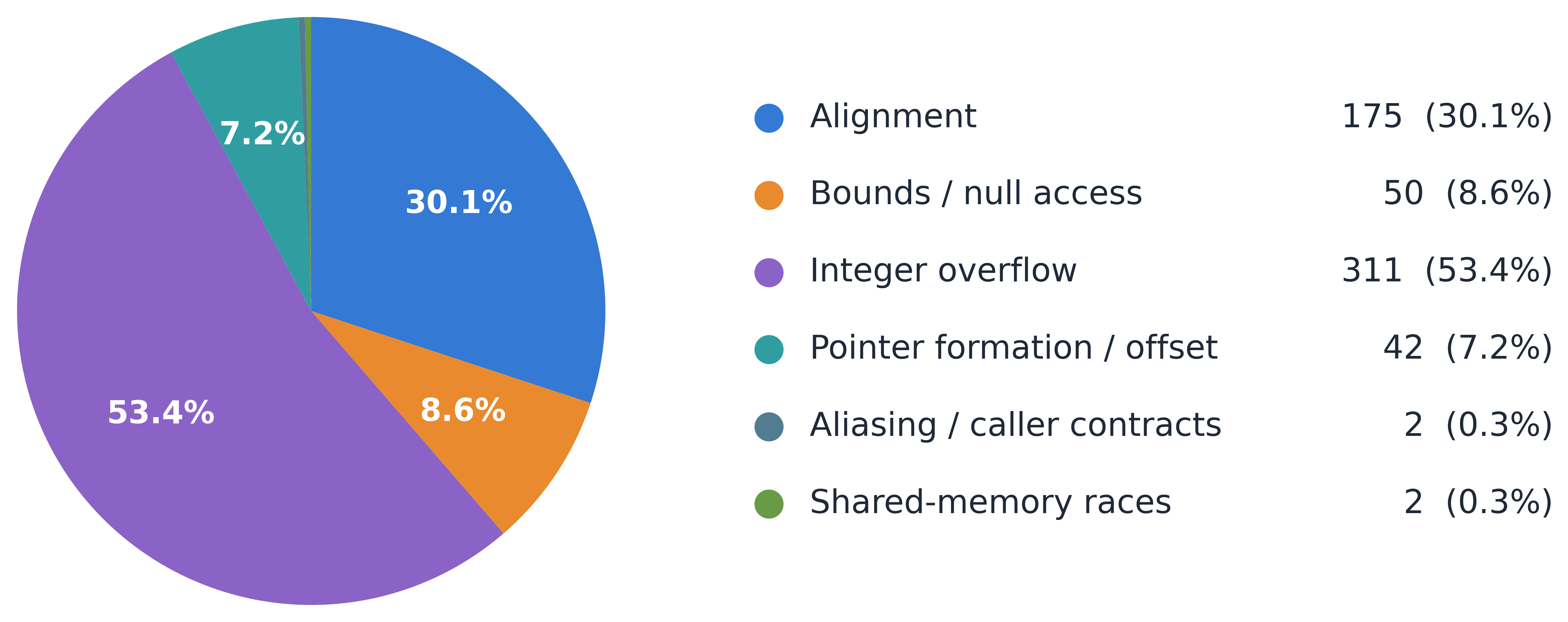}}
\caption{\textbf{\mytool{} bug classification across SOL-ExecBench, Optimized CUDA, and KernelBench suites}. Each of the 582 instances with potential findings contributes to one primary category.}
\label{bug-class}
\end{figure}

\textbf{Classification of Bugs}. A deeper dive into the 582 potential bug instances identified by \mytool{} reveals that integer overflow (311; 53.4\%) and misaligned accesses (175; 30.1\%) dominate, together accounting for 83.5\% of primary findings (Fig. \ref{bug-class}). The remaining findings concern null accesses, invalid pointer formations and memory races. These overflow and alignment issues mostly occur due to unchecked arithmetic such as index calculations, dimension products leading to overflow before division/widening, or insufficient packed-access assumptions whose alignment requirements are not guaranteed by the underlying storage. This highlights a recurring gap between the inputs accepted by wrappers and the assumptions required by optimized implementations in modern agentic code-generation systems.

\begin{lstlisting}[style=cuda,
    caption={64-bit cast applied too late: MaxPool2D},
    label={lst:cuda-max}]
__global__ void maxpool2d_kernel(..) {
const int64_t total = (int64_t)B * C * OH * OW;
// ...
const float* xc = x + ((int64_t)(b * C + c)) * H * W;
}
\end{lstlisting}

For instance, Listing \ref{lst:cuda-max} shows a subtle example where a 64-bit cast masks an overflow risk: the pointer-offset expression \texttt{b * C + c} is evaluated in 32-bit arithmetic before conversion, potentially overflowing for large tensor dimensions. For this finding, \mytool{} reports the configuration $b = 65536$, $C = 32768$, and $c = 0$, for which the intermediate product exceeds the signed 32-bit range.


%

\begin{table}[t]
\centering
\caption{\mytool{} verification outcomes across 740 instances
containing 1,370 CUDA kernel definitions.
Verified$^*$ denotes successful verification under stated
kernel/library preconditions.}
\label{tab:verification-outcomes}
\footnotesize
\setlength{\tabcolsep}{2pt}
\renewcommand{\arraystretch}{1.15}
\resizebox{\columnwidth}{!}{%
\begin{tabular}{@{}l *{4}{c}@{}}
\toprule
\multicolumn{5}{@{}c}{\textbf{SOL-ExecBench}} \\
\addlinespace[2pt]
Domain & Instances & Verified$^*$ &
\shortstack{Potential\\Bug} & Inconclusive \\
\midrule
L1 (Single Operations) & 25 & 6 & 14 & 5 \\
L2 (Fused Operations)  & 20 & 2 & 14 & 4 \\
Quantization          & 3  & 1 & 1  & 1 \\
FlashInfer-Bench       & 16 & 2 & 10 & 4 \\
\textbf{Subtotal} & \textbf{64} & \textbf{11} &
\textbf{39} & \textbf{14} \\
\midrule
\multicolumn{5}{@{}c}{\textbf{Optimized CUDA}} \\
\addlinespace[2pt]
Workload & Instances & Verified &
\shortstack{Potential\\Bug} & Inconclusive \\
\midrule
Attention / transformers & 9 & 9 & 0 & 0 \\
Norm. / modulation       & 8 & 5 & 3 & 0 \\
Convolution              & 2 & 2 & 0 & 0 \\
MLP / MoE                & 4 & 4 & 0 & 0 \\
State-space models       & 3 & 3 & 0 & 0 \\
GEMM                     & 1 & 1 & 0 & 0 \\
Quantum simulation       & 1 & 1 & 0 & 0 \\
\textbf{Subtotal} & \textbf{28} & \textbf{25} &
\textbf{3} & \textbf{0} \\
\midrule
\multicolumn{5}{@{}c}{\textbf{KernelBench}} \\
\addlinespace[2pt]
Workload & Instances & Verified &
\shortstack{Potential\\Bug} & Inconclusive \\
\midrule
GEMM / GEMV        & 103 & 8  & 64  & 31 \\
Activations        & 112 & 19 & 92  & 1  \\
Norm. / softmax    & 167 & 6  & 151 & 10 \\
Pooling            & 54  & 0  & 54  & 0  \\
Reductions / scans & 108 & 9  & 95  & 4  \\
Fused GEMM         & 104 & 6  & 84  & 14 \\
\textbf{Subtotal} & \textbf{648} & \textbf{48} &
\textbf{540} & \textbf{60} \\
\bottomrule
\end{tabular}%
}
\end{table}

\subsection{\mytool{}'s Overheads \& Cost Analysis}

\textbf{Timing Overheads}. Table \ref{tab:overheads} breaks down \mytool{}’s analysis overhead, showing that annotation and verification dominate preprocessing. Verification accounts for more than half of their combined time—57\% on SOL-ExecBench and 52\% on KernelBench. This is expected because each annotation-refinement iteration requires translating the annotated CUDA and its obligations through VerCors.

Note that \mytool{} requires more turns on KernelBench cases than on SOL-ExecBench (5.35 versus 3.92), despite the latter having approximately 4.9$\times$ as many LOC per case. This is largely driven by 60 unproved KernelBench cases that exhaust the 12-turn repair budget.

\begin{table*}[t]
\centering
\caption{Timing analysis of different phases of \mytool{}'s agentic workflow}
\label{tab:overheads}
\resizebox{\textwidth}{!}{%
\begin{tabular}{l*{8}{r}}
\toprule
& \multicolumn{2}{c}{Avg. preprocessing (sec)}
& \multicolumn{2}{c}{Avg. verification (sec)}
& \multicolumn{2}{c}{Avg. annotation (sec)}
& \multicolumn{2}{c}{Avg. turns (limit = 12)} \\
\cmidrule(lr){2-3}
\cmidrule(lr){4-5}
\cmidrule(lr){6-7}
\cmidrule(lr){8-9}
& Per kernel & Per Instance
& Per kernel & Per Instance
& Per kernel & Per Instance
& Per kernel & Per Instance \\
\midrule
SOL-ExecBench
& 0.78 & 2.76
& 111.12 & 393.38
& 82.21 & 291.01
& 1.11 & 3.92 \\
KernelBench
& 0.22 & 0.34
& 456.56 & 740.16
& 424.81 & 688.68
& 3.30 & 5.35 \\
\bottomrule
\end{tabular}%
}
\end{table*}

\textbf{Tokens \& Estimated Cost}. Table \ref{tab:token-cost} estimates an average model cost of \$3.58 per instance and \$2.26 per kernel, including recorded annotation and repair requests. This suggests relatively modest inference costs for establishing safety in production-grade kernels. 

\begin{table}[t]
\centering
\caption{\mytool{}'s estimated average token usage and model cost, based on estimated input tokens, recorded output tokens, and GPT-6 Astra Standard API pricing.}
\label{tab:token-cost}
\footnotesize
\setlength{\tabcolsep}{3pt}
\renewcommand{\arraystretch}{1.15}
\begin{tabularx}{\columnwidth}{
    @{}*{4}{>{\centering\arraybackslash}X}@{}
}
\toprule
Avg tokens/ instance &
Avg tokens/ kernel &
Avg price/ instance &
Avg price/ kernel \\
\midrule
$\approx 290{,}114$ &
$\approx 183{,}011$ &
$\approx \$3.58$ &
$\approx \$2.26$ \\
\bottomrule
\end{tabularx}
\end{table}

\subsection{Comparison with state-of-the-art tools}

To evaluate \mytool{}'s effectiveness in detecting memory and thread safety violations, we compare it against state-of-the-art tools on previously published benchmark suites.

\textbf{Comparison with State-of-the-art Memory Safety Tools}:~\autoref{tab:safety-comparison} compares the memory safety coverage provided by \mytool{} against four state-of-the-art GPU dynamic memory safety tools including NVIDIA's Compute Sanitizer memcheck \cite{nvidiaComputeSanitizer}, GMOD \cite{di2018gmod}, GPUShield \cite{lee2022gpushield}, and cuCatch \cite{ibnziad2023cucatch} using the cuCatch benchmark suite \cite{ibnziad2023cucatch}. Our experiments show that \mytool{} identified 100\% of the memory safety bugs in the benchmark suite compared to 35.7\%, 14.3\%, 42.9\%, and 71.4\% for Compute Sanitizer, GMOD, GPUShield, and cuCatch respectively. The lower coverage rates in these tools reflect limitations in detecting different classes of memory errors: Compute Sanitizer missed non-adjacent buffer overflows, while cuCatch, despite addressing some of these cases, still missed intra-allocation out-of-bounds accesses. On the other hand, \mytool{} not only correctly identified all memory safety errors in the benchmark but also identified a bug in the benchmark suite that makes a potential out-of-bounds access unreachable.

\begin{table*}[t]
\centering
\caption{\mytool{}'s memory-safety comparison with state-of-the-art memory safety tools on cuCatch's benchmark suite. \mytool{} achieved 100\% detection rate; * indicates we used these tools' published numbers \cite{ibnziad2023cucatch}.}
\label{tab:safety-comparison}
\small
\setlength{\tabcolsep}{5pt}
\renewcommand{\arraystretch}{1.2}
\begin{tabularx}{\textwidth}{
    @{}l
    >{\raggedright\arraybackslash}X
    *{5}{c}@{}
}
\toprule
\textbf{Category} & \textbf{Description} &
\shortstack{\textbf{Compute}\\\textbf{Sanitizer}} &
\textbf{GMOD}$^*$ &
\textbf{GPUShield}$^*$ &
\textbf{cuCatch} &
\textbf{\mytool{}} \\
\midrule
Temporal safety
    & Lifetime and deallocation errors.
    & 8/12 & 4/12 & 4/12 & 10/12 & 12/12 \\
Spatial heap
    & OOB accesses to heap objects.
    & 4/16 & 4/16 & 8/16 & 8/16 & 16/16 \\
Spatial stack
    & OOB accesses to stack objects.
    & 4/16 & 0/16 & 12/16 & 12/16 & 16/16 \\
Spatial shared memory
    & OOB accesses to shared objects.
    & 4/12 & 0/12 & 0/12 & 10/12 & 12/12 \\
\midrule
\textbf{Detection rate} & &
\shortstack{\textbf{20/56}\\(35.7\%)} &
\shortstack{\textbf{8/56}\\(14.3\%)} &
\shortstack{\textbf{24/56}\\(42.9\%)} &
\shortstack{\textbf{40/56}\\(71.4\%)} &
\shortstack{\textbf{56/56}\\(100\%)} \\
\bottomrule
\end{tabularx}
\end{table*}

\textbf{Comparison with State-of-the-art Race Detectors}: We compared \mytool{}'s race detection coverage against state-of-the-art dynamic race detectors for GPUs, including iGUARD \cite{kamath2021iguard}, NVIDIA's Compute Sanitizer Racecheck \cite{nvidiaComputeSanitizer}, and the more recent SuperCollider tool \cite{stephenson2026supercollider}.~\autoref{tab:cuhadron-comparison} shows the results of our study on cuHadron \cite{stephenson2026supercollider}, a GPU race detection benchmark designed to evaluate the coverage of GPU race detection tools on modern GPU features, including asynchronous memory copies.

Unlike the dynamic detectors, which execute the benchmark programs to observe races, \mytool{} analyzes kernels statically. It detects races in all 14 tests within its kernel-level analysis scope and correctly verifies all four race-free control tests. The remaining nine race tests involve interactions across streams, CPUs and GPUs, or multiple GPUs, which exceed its current analysis scope.\footnote{This reflects the current implementation's supported scope, rather than an inherent limitation of the technique.} Nevertheless, \mytool{} identifies additional potential memory-safety issues: signed integer overflows in three of these tests, demonstrating useful diagnostic coverage even where the intended race cannot be analyzed.


Across all 23 race-positive tests, \mytool{} shows 61\% coverage on the entire test suite, as compared to 4\%, 39\%, and 87\% for iGUARD, Compute Sanitizer and SuperCollider respectively, and 100\% coverage on in-scope tests with zero false positives. Within its scope, \mytool{} detects all 14 races, including those involving asynchronous memory copies, without executing the benchmark programs, and produces no false positives on the four race-free controls.

\begin{table*}[t]
\centering
\small
\setlength{\tabcolsep}{4pt}
\caption{\mytool{} compared with state-of-the-art race detectors on cuHadron suite.
\emph{In-scope} races occur between threads of a single kernel launch, including
asynchronous copies issued by that kernel; \emph{out-of-scope} races involve another
kernel, a host-issued copy, another GPU or the CPU. Each cell gives planted races
detected out of races in the category. \WR{} counts tests where \mytool{} detected a different, non-race, memory safety error (signed integer
overflow) in the benchmark. For the false-positive controls, the cell gives race-free tests
incorrectly flagged. * indicates we used these tools' published numbers \cite{stephenson2026supercollider}.}
\label{tab:cuhadron-comparison}
\begin{tabular}{l l p{4.6cm} cccc}
\toprule
\textbf{Scope} & \textbf{Category} & \textbf{Description} & \textbf{iGUARD$^*$} & \makecell{\textbf{Compute}\\\textbf{Sanitizer}} & \makecell{\textbf{SuperCollider}} & \makecell{\textbf{KernelArmor}} \\
\midrule
\multirow{5}{*}{In scope}
  & Inter-subwarp    & Divergent threads within a warp                  & 0/2 & 2/2 & 2/2 & 2/2 \\
  & Async memcpy     & \texttt{cp.async}(\texttt{.bulk}) accesses & 0/9 & 5/9 & 8/9 & 9/9 \\
  & CGA & Distributed SMEM Accesses            & 0/2 & 2/2 & 2/2 & 2/2 \\
  & Inter-wave       & Thread blocks in different waves                 & 1/1 & 0/1 & 1/1 & 1/1 \\
  & \multicolumn{2}{l}{\textit{Subtotal}}                               & 1/14 & 9/14 & 13/14 & \textbf{14/14} \\
\midrule
\multirow{5}{*}{Out of scope}
  & Inter-kernel     & Kernels on different streams          & 0/2 & 0/2 & 2/2 & 0/2 \\
  & cudaMemcpyAsync  & Async copies vs.\ kernels         & 0/2 & 0/2 & 1/2 & 0/2 (+1\,\WR) \\
  & Inter-GPU        & Multiple GPUs on unified memory                  & 0/2 & 0/2 & 2/2 & 0/2 \\
  & CPU--GPU         & CPU and GPU on unified memory                    & 0/3 & 0/3 & 2/3 & 0/3 (+2\,\WR) \\
  & \multicolumn{2}{l}{\textit{Subtotal}}                               & 0/9 & 0/9 & 7/9 & 0/9 (+3\,\WR) \\
\midrule
\multicolumn{3}{l}{\textbf{Total races}}                                & 1/23 & 9/23 & 20/23 & 14/23 (+3\,\WR) \\
\midrule
\multicolumn{3}{l}{False positives (race-free: threads access different bytes of one word)} & 2/4 & 0/4 & 0/4 & 0/4 \\
\bottomrule
\end{tabular}
\end{table*}

\textbf{Comparison with other agentic methods}: Finally, we compare \mytool{} with ProofWright \cite{chatterjee2025proofwright}, a closely related
agentic formal verification framework that also establishes memory and thread safety for CUDA kernels. We evaluated \mytool{} on the 100
Claude-4-Sonnet legacy implementations of KernelBench Level~1 problems
used in ProofWright's evaluation. These implementations
primarily rely on simple CUDA constructs, including shared
memory accesses, affine thread and block indexing and block-level
synchronization.

Table~\ref{tab:agentic-comparison} summarizes the results of our study.
On the legacy KernelBench corpus, \mytool{} establishes memory- and
thread-safety guarantees for 89\% of the kernels, compared with ProofWright's 74\%. Additionally, \mytool{} identifies 10 potential overflow bugs supported by
concrete counter-examples. In contrast, none of the 1,370 kernels in \mytool{}'s expanded benchmark corpus is supported by ProofWright, as they rely on advanced CUDA features that fall outside its supported verification scope.

\begin{table*}[t]
\centering
\caption{Verification outcomes for
ProofWright and \mytool{} on legacy and expanded CUDA benchmark suites.}
\label{tab:agentic-comparison}
\small
\setlength{\tabcolsep}{5pt}
\renewcommand{\arraystretch}{1.2}
\begin{tabularx}{\textwidth}{
    @{}>{\raggedright\arraybackslash}X l rrrr@{}
}
\toprule
\textbf{Benchmark corpus} & \textbf{Tool} & \textbf{Verified} & \textbf{Potential bug}
& \textbf{Inconclusive} & \textbf{Unsupported} \\
\midrule
Legacy KernelBench Problems (100 kernels)
& ProofWright & 74\% & N/A & 24\% & 2\% \\
& \mytool{} & 89\% & 10\% & 1\% & 0\% \\
\midrule
SOL-ExecBench + KernelBench + optimized CUDA
& ProofWright & 0\% & N/A & 0\% & 100\% \\
(1,370 kernels) & \mytool{} & 11.4\% & 78.6\% & 6.6\% & 3.4\% \\
\bottomrule
\end{tabularx}
\end{table*}

\subsection{Ablation Study: Impact of \mytool{}'s Verification Handbook}

We evaluate the contribution of \mytool{}'s Verification Handbook
using six randomly selected cases from our benchmark suites. For each case, we compare a handbook-enabled run against an ablated
run that removes the handbook from the model's reference context.
Both conditions use the same source input, model, verifier settings,
and budget of 16 model turns. 

Table~\ref{tab:annotation-guide-ablation} shows that \mytool{} verifies three of six cases with the handbook, compared with two without it. Among cases verified in both conditions, removing the handbook increases model turns from 6 to 12 for Irregular Matmul and from 7 to 8 for ArgminReduction. These experiments show that the Verification Handbook is essential both for convergence and for correctness of \mytool{}.


\begin{table}[t]
\centering
\caption{Verification handbook ablation study for \mytool{}
on six randomly selected kernels, with a limit of 16 turns
per kernel.}
\label{tab:annotation-guide-ablation}
\footnotesize
\setlength{\tabcolsep}{3pt}
\renewcommand{\arraystretch}{1.15}
\begin{tabularx}{\columnwidth}{
    @{}>{\raggedright\arraybackslash}X
    >{\raggedright\arraybackslash}p{0.18\columnwidth}
    r
    >{\raggedright\arraybackslash}p{0.18\columnwidth}
    r@{}
}
\toprule
& \multicolumn{2}{c}{\textbf{With handbook}}
& \multicolumn{2}{c}{\textbf{Without handbook}} \\
\cmidrule(lr){2-3}\cmidrule(l){4-5}
\textbf{Task} & Outcome & Turns & Outcome & Turns \\
\midrule
LatentKV
    & Potential bug & \textbf{3} & Exhausted & 16 \\
\addlinespace[4pt]
SOL RMSNorm\allowbreak H1536
    & Potential bug & \textbf{7} & Potential bug & 8 \\
\addlinespace[4pt]
KB Upper Triangular Matmul
    & Potential bug & \textbf{3} & Exhausted & 16 \\
\addlinespace[4pt]
KB Irregular Matmul
    & Verified & \textbf{6} & Verified & 12 \\
\addlinespace[4pt]
KB Matmul\allowbreak LargeK
    & Verified & \textbf{7} & Potential bug & 13 \\
\addlinespace[4pt]
KB Argmin\allowbreak Reduction
    & Verified & \textbf{7} & Verified & 8 \\
\midrule
\textbf{Verified cases}
    & \multicolumn{2}{c}{\textbf{3/6}}
    & \multicolumn{2}{c}{2/6} \\
\bottomrule
\end{tabularx}
\end{table}

\subsection{Integrating \mytool{} into an Agentic Optimization Loop}
\label{sec:eval-loop}

\begin{figure}[t]
\centering
\begin{tikzpicture}[
    box/.style={draw, rounded corners=2pt, align=center, minimum height=10mm, text width=13mm,
        font=\scriptsize, line width=0.5pt, inner sep=2pt},
    lbl/.style={font=\tiny, align=center, inner sep=1pt},
    arr/.style={-{Stealth[length=1.6mm]}, line width=0.5pt}]
\node[box, fill=agentfill, draw=agentdraw] (agent) {\textbf{Agent}\\proposes a kernel};
\node[box, fill=evalfill, draw=evaldraw, right=8mm of agent] (eval) {\textbf{Evaluate}\\compile, check, time};
\node[box, fill=kgfill, draw=kgdraw, text width=15mm, right=5mm of eval] (kg) {\textbf{\mytool{}}\\verifies a new best};
\node[box, fill=bestfill, draw=bestdraw, right=9mm of kg] (best) {\textbf{Accept}\\as new best};
\draw[arr] (agent) -- node[lbl, above] {kernel} (eval);
\draw[arr] (eval) -- (kg);
\draw[arr] (kg) -- node[lbl, above] {verified} (best);
\draw[arr, dashed] (eval.north) -- ++(0,3mm) -| node[lbl, above, pos=0.25] {unconstrained arm} (best.north);
\draw[arr] (eval.south) -- ++(0,-3mm) -| node[lbl, below, pos=0.25] {correctness, time} ([xshift=3mm]agent.south);
\draw[arr] (kg.south) -- ++(0,-8mm) -| node[lbl, below, pos=0.25] {verdict: failing obligation, location, or explanation} ([xshift=-3mm]agent.south);
\end{tikzpicture}
\caption{The gated optimization loop. Each round, the agent's kernel is compiled, checked against the PyTorch reference, and timed. A correct kernel that would become the new best is accepted only if \mytool{} verifies it, and the verdict returns to the agent as feedback. The unconstrained arm accepts it directly.}
\label{fig:loop}
\end{figure}
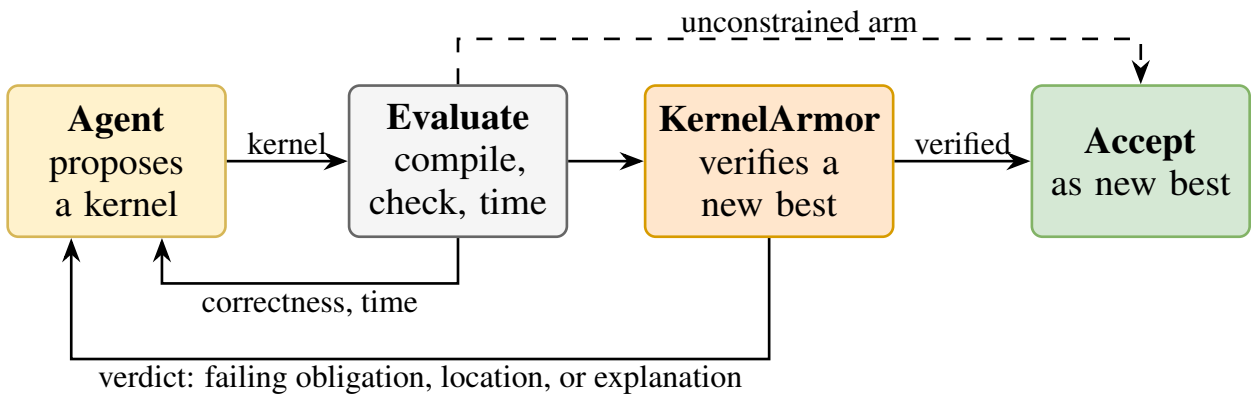

\textbf{Loop setup}: We place \mytool{} inside an agentic CUDA optimization loop (Figure~\ref{fig:loop}) in which a coding agent (Claude Opus 4.8) replaces a PyTorch operator with a CUDA kernel and improves it over eight rounds. We compare two arms on 11 KernelBench~\cite{ouyang2025kernelbench} tasks (Table~\ref{tab:inloop}), with seven independent runs (trajectories) per task and arm. Each trajectory starts from scratch, and trajectories differ only through the agent's sampling. The \textit{unconstrained} arm keeps the fastest correct kernel. In the \textit{gated} arm, every correct candidate that would become the new best is first verified by \mytool{}, and only a \textit{verified} kernel is accepted. The agent receives \mytool{}'s verdict as feedback: the failing obligation and its source location for a potential bug, the fact that no proof was found for an inconclusive result, and the verifier's explanation when a construct is unsupported. The gated prompt asks for kernels that are safe on every input the host wrapper accepts. Each trajectory has a 6-hour budget. Kernels run on an NVIDIA A40 (48\,GB, driver 590.48.01, CUDA 13.1) with clocks locked at 1410\,MHz. We select each trajectory's best kernel by its in-loop time and report its speedup over the PyTorch reference from a separate benchmark run.

\begin{table}[t]
\centering
\small
\caption{Median best-kernel speedup over PyTorch across seven trajectories per arm. \textit{Ratio} is the geometric mean of gated over unconstrained speedups. \textit{Safe} counts unconstrained best kernels that \mytool{} verifies. All gated best kernels are verified.}
\label{tab:inloop}
\resizebox{\columnwidth}{!}{%
\begin{tabular}{lrrrr}
\toprule
Task & Unconstr. & Gated & Ratio & Safe \\
\midrule
ReLU & 1.04$\times$ & 1.04$\times$ & 0.99 & 6/7 \\
LogSoftmax & 1.28$\times$ & 1.27$\times$ & 0.90 & 0/7 \\
GroupNorm & 1.00$\times$ & 1.00$\times$ & 0.99 & 0/7 \\
LayerNorm & 8.31$\times$ & 8.28$\times$ & 0.98 & 0/7 \\
MaxPool 2D & 2.81$\times$ & 2.62$\times$ & 0.89 & 0/7 \\
Matrix-vector & 1.54$\times$ & 1.58$\times$ & 1.00 & 0/7 \\
Argmax & 0.91$\times$ & 0.89$\times$ & 0.96 & 0/7 \\
GELU & 9.30$\times$ & 9.30$\times$ & 1.00 & 5/7 \\
cumsum & 1.18$\times$ & 1.17$\times$ & 0.99 & 0/7 \\
Gemm+GroupNorm+Hardtanh & 18.86$\times$ & 21.50$\times$ & 1.13 & 1/7 \\
Gemm+Add+ReLU & 21.16$\times$ & 21.53$\times$ & 1.06 & 0/7 \\
\midrule
All (77 trajectories) & & & 0.99 & 12/77 \\
\bottomrule
\end{tabular}%
}
\end{table}

\textbf{Speed and safety}: Across all 77 trajectories, the gated arm's geometric-mean speedup is 0.99$\times$ that of the unconstrained arm, and every accepted gated kernel is verified (Table~\ref{tab:inloop}). Per task, the gated median is within 7\% of the unconstrained median, and the largest gap is MaxPool. The unconstrained arm, by contrast, ends with a verified kernel in only 12 of its 77 trajectories, all on ReLU, GELU, and one Gemm+GroupNorm+Hardtanh trajectory. The findings on its best kernels are signed-integer overflow at sizes near \texttt{INT\_MAX} (for example, \texttt{N*C} in GroupNorm and launch-grid ceiling divisions in MaxPool), \texttt{float4} misalignment on offset tensor views, missing output-size checks, pointers formed outside their allocation, and an out-of-bounds shared-memory write at batch size 257 in LayerNorm. None of them triggers on the benchmark shapes, so testing on those shapes cannot expose them.

\textbf{Post-hoc filtering}: A fair question is whether it suffices to run the unconstrained loop and keep only its verified kernels afterwards. Doing so leaves a kernel in just 31 of 77 trajectories, and these are often slow early-round kernels: 0.16$\times$ for cumsum and 1.74$\times$ for LayerNorm against 8.3$\times$ in the loop. With \mytool{} in the loop, the agent instead fixes the reported defects and keeps optimizing. The fixes we observe, such as 64-bit or bounded ceiling divisions, a scalar fallback for misaligned \texttt{float4} views, and an overlap check between input and output, cost no measurable speed. For example, \mytool{} rejected a 6.4$\times$ Gemm+GroupNorm+Hardtanh kernel because \texttt{B * G} overflows a signed integer at $B = 2^{27}$. The agent's next round kept the cuBLAS-based design and was verified. Verification therefore pays off inside the loop, where its feedback steers the search, not after it.

\textbf{Verifier capacity}: Gating is not free, however. The gated arm loses speed mainly when \mytool{} cannot decide a fast candidate. In 13 of the 77 trajectories, the gate rejected a candidate more than 1\% faster than the kernel the trajectory ended with: in 7 of them the fastest such candidate was inconclusive, in 4 it had a potential bug, and in 2 it used an unsupported library configuration. Verification also costs time: a single \mytool{} run takes a median of 12 minutes, and the median gated trajectory spends 1.45\,h in \mytool{} across all its rounds (at most 5.3\,h). Across the 341 verification runs, 243 ended \textit{verified}, 63 \textit{potential bug}, 28 \textit{inconclusive}, 5 unsupported, and 2 errored. In exchange for 1\% of geometric-mean speed, every gated kernel comes with a memory- and thread-safety proof.

\textbf{Library contracts}: In our initial experiments on the two GEMM tasks, we noticed that the gated agent incorrectly abandoned cuBLAS, assuming the verifier could not reason about library calls. It reached only 2.2--2.6$\times$, against 19--21$\times$ unconstrained. We therefore added one prompt sentence stating that cuBLAS calls are checked against library contracts (\S\ref{sec:proofwright}). All 14 trajectories then ended with a verified kernel that calls cuBLAS for the GEMM and fuses the epilogue, at 21.5$\times$. Similarly, when the feedback for an unsupported construct named only the blocker class, the agent abandoned a 6.6$\times$ cuBLAS kernel. Once \mytool{} explained the unsupported call, the agent switched to a supported one instead. An in-the-loop verifier must therefore tell the agent what it can check and why it rejects a candidate, not only whether the candidate passed.

\section{Related Work}
\label{sec:related}

\paragraph{LLM-assisted deductive verification.}
AutoVerus and dafny-annotator use language models to supply proof annotations and verifier feedback to guide subsequent attempts~\cite{yang2025autoverus,poesia2024dafnyannotator}.
ESBMC-AI uses bounded-model-checking counterexamples to guide repairs to vulnerable C programs~\cite{tihanyi2023esbmcai}.
ProofWright applies this approach to CUDA, using a VerCors knowledge base and an evolving annotation guide for safety proofs, together with Rocq-based reasoning for selected semantic checks~\cite{chatterjee2025proofwright}.
\mytool{} builds on this agentic verification methodology and addresses the CUDA-modeling limitations that constrain its application to optimized kernels.
Its contributions include the extended safety model, annotation-only integrity checks against the preprocessed program, and diagnostics that distinguish validated safety findings from unresolved proofs and unsupported constructs.
The annotation guard preserves the proof target; the resulting safety guarantee remains conditional on the stated contracts and modeling assumptions.

\paragraph{Verified GPU programming and numerical correctness.}
Kuiper uses F*/Pulse to construct safe, functionally verified GPU programs~\cite{martinez2026kuiper}; KuiperBench applies this approach to agent-written implementations of all 100 KernelBench Level~1 tasks~\cite{martinez2026kuiperbench}.
Browne et al. combine Float32 equivalence proofs in Lean with VerCors safety verification~\cite{browne2026gpuverification}.
\mytool{} focuses on memory and thread safety of existing CUDA implementations under explicit contracts, leaving numerical equivalence as a complementary objective.

\paragraph{Dynamic safety checking and fuzzing.}
Compute Sanitizer, cuCatch, iGUARD, and SuperCollider detect memory errors or races at runtime~\cite{nvidiaComputeSanitizer,ibnziad2023cucatch,kamath2021iguard,stephenson2026supercollider}.
cuFuzz expands input exploration through host/device coverage, but its coverage signatures do not distinguish thread interleavings~\cite{ibnziad2026cufuzz}.
These tools depend on exercised inputs and schedules; \mytool{} complements them with deductive safety guarantees under explicit contracts and its supported CUDA model.

\paragraph{Static GPU safety analysis.}
GPUVerify reduces race- and barrier-divergence checking to sequential verification~\cite{betts2012gpuverify}; its documented default configuration does not check array bounds~\cite{gpuverifyLimitations}.
Faial infers access-memory protocols for compositional race analysis without user-supplied loop invariants~\cite{cogumbreiro2021faial}.
GKLEE uses concolic execution for verification and test generation~\cite{li2012gklee}, and ESBMC-GPU explores CUDA behaviors through SMT-based context-bounded model checking~\cite{monteiro2018esbmcgpu}.
More recently, SCuBA relates host-side allocation to device-side accesses and tracks logical allocation partitions to detect input-dependent and intra-allocation bounds errors~\cite{ghosh2026elusive}.
\mytool{} shares the goal of detecting errors beyond the tested input distribution and extends VerCors' separation-logic approach~\cite{blom2014gpgpu,blom2017vercors}, combining bounds and alignment checks with permission-based concurrency reasoning.

\section{Conclusion}
\label{sec:conc}
This paper presents \mytool{}, which combines agent-generated annotations with an extended SMT Verifier to establish memory and thread safety for optimized CUDA kernels under explicit contracts. It outperforms the evaluated state-of-the-art tools on cuCatch and in-scope cuHadron tests. Integrated into an agentic optimization loop, it produces verified kernels in all trajectories while retaining 99\% of the unconstrained geometric-mean speedup.


\bibliography{example_paper}
\bibliographystyle{mlsys2025}

\appendix


\end{document}